\documentclass[aps,prl,twocolumn,citeautoscript,groupedaddress,amsmath,amssymb]{revtex4}

\usepackage{graphics}
\graphicspath{{fig/}}

\usepackage{dcolumn}
\usepackage{bm}
\usepackage{units}

\usepackage[final]{LLNdef}

\def\Han{H_{\mathrm{ann}}}
\def\Hn{H_{\mathrm{nuc}}}
\def\Hsw{H_{\mathrm{swi}}}

\begin{document}


\title{Angular-dependence of magnetization switching for a multi-domain dot: experiment and simulation}

\author{O. Fruchart}
   \email{Olivier.Fruchart@grenoble.cnrs.fr}
\author{J.-C. Toussaint}
\author{P.-O. Jubert}
 \altaffiliation[Present address: ]{IBM Research - Zurich Research Laboratory, 8803 R{\"u}schlikon, Switzerland}
\author{W. Wernsdorfer}
  \affiliation{Laboratoire Louis N{\'e}el, CNRS, BP166, F-38042 Grenoble Cedex 9, France}
\author{R. Hertel}
  \affiliation{Max Plank Institut f{\"u}r Mikrostrukturphysik, D-06120, Halle, Germany}
\author{D. Mailly}
  \affiliation{Laboratoire de Photonique et de Nanostructures, CNRS, Route de Nozay,  F-91460 Marcoussis, France}
\author{J. Kirschner}
  \affiliation{Max Plank Institut f{\"u}r Mikrostrukturphysik, D-06120, Halle, Germany}

\date{\today}

\begin{abstract}
We have measured the in-plane angular variation of nucleation and annihilation fields of a
multi-domain magnetic single dot with a \microsquid. The dots are Fe/Mo(110) self-assembled in
UHV, with sub-micron size and a hexagonal shape. The angular variations were quantitatively
reproduced by micromagnetic simulations. Discontinuities in the variations are observed, and shown
to result from bifurcations related to the interplay of the non-uniform magnetization state with
the shape of the dot.
\end{abstract}

\pacs{Valid PACS appear here}
\maketitle

\vskip 0.5in

\def\CR{CR}
\def\SD{SD}

Coherent rotation of magnetization is the simplest model of magnetization reversal, proposed by
Stoner and Wohlfarth in 1948\cite{bib-STO48}. Coherent rotation predicts the value of the
switching field $\Hsw$ of a single-domain system as a function of the direction of the external
field~$\Hext$. For a two-dimensional system with uniaxial anisotropy the polar plot
$\Hsw(\varphi)$ falls on the well-known astroid\cite{bib-SLO56}. The full experimental proof for
coherent rotation was given only recently, when nanoparticles of high quality and of size small
enough to roughly satisfy the hypothesis of uniform magnetization could be investigated
individually\cite{bib-WER97,bib-BON99}. Starting from this proof, it is now a challenge to
understand magnetization reversal in increasingly large (and thus complex) systems. The simplest
ingredient to add to coherent rotation is to allow minor deviations from strictly homogeneous
magnetization. The consequences on magnetization processes were addressed by numerical
micromagnetics\cite{bib-SCH88}, investigated analytically\cite{bib-COW98b} and checked
experimentally\cite{bib-COW98c}. The next step is now to tackle quantitatively more strongly
non-uniform systems, those that may display magnetic domains and domain walls\cite{bib-HUB98b}. In
such systems a switching field $\Hsw$ is not the signature of the full reversal of magnetization,
but instead reflects events like nucleation, propagation and annihilation\cite{bib-WER96c}. Few
and only partial experimental\cite{bib-WER96c} and numerical\cite{bib-WEI02} reports are found on
this issue. A more detailed study would open the door to understanding microscopic details of
magnetization reversal processes in macroscopic materials. In this Letter we present such a study
in a model system: sub-micrometer Fe faceted dots self-assembled in UHV, that have a high
structural quality and display simple multi-domain states\cite{bib-FRU01b,bib-FRU03c}. The angular
dependence of the $\Hsw$'s of a single dot was studied with the \microsquid\
technique\cite{bib-WER01}. This can be seen as the first experimental generalization of astroids
for an individual multi-domain system. A striking feature is the occurrence of
discontinuities~(hereafter named jumps) in $\Hsw(\varphi)$ plots. These jumps were reproduced and
understood with the help of numerical micromagnetism. They result from bifurcations, related to
the interplay of the non-uniform magnetization with the shape of the dot. This also shows that a
complex $\Hsw$ behavior does not necessarily result from defects, whose role may have been
overestimated in the literature of magnetic dots made by lithography.

\begin{figure}[h]
  \includegraphics{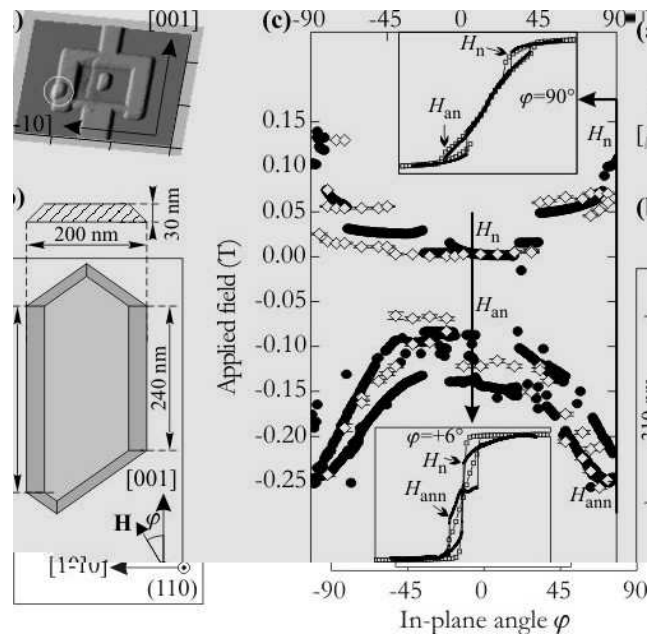}%
  \caption{\label{fig-afm-loop-polarplots}\dataref{none}(a)~AFM picture of the
  \microsquid. The dot strongly coupled to the \microsquid is indicated by a circle.
   (b) real-scale top view of the dot (c)~Plot of $\Hn$~(positive) and $\Han$~(negative) for experiments~(full symbols)
  and simulations~(open symbols).
  Insets~: experimental and simulated loops for two angles.}
\end{figure}

The Fe(110) epitaxial dots were fabricated with pulsed laser deposition in ultra-high vacuum by
self-assembly on $\mathrm{Mo}(110)[\thicknm{8}]/\saphir(11\overline20)$. The dots display the
shape of ingots with atomically-flat facets, bulk lattice parameter and bulk cubic
magneto-crystalline anisotropy $K_1$ favoring $<100>$ axes, however of magnitude much smaller than
$\frac12\muZero\Ms^2$\cite{bib-FRU01b}. The inter-dot dipolar fields are negligible with respect
to~$\Hsw$. The remanent state consists of flux-closure domains, resulting from demagnetizing
fields within each dot\cite{bib-FRU03c}. Such domains can occur due to size of the dots being well
above the exchange length~$\ExchangeLength$\cite{bib-FRE57,bib-KRO62}. The in-plane $\Hsw$'s of a
single Fe dot were measured below $\tempK4$ using the \microsquid\ technique\cite{bib-WER01}. For
these measurements, the dots were covered in UHV by Mo$[\thicknm{2}]$, followed by
$\mathrm{Al}[\thicknm2]$~(then 12 hours air-oxidized), and a $\mathrm{Si}[\thicknm2]\backslash
\mathrm{Nb}[\thicknm{15}]\backslash \mathrm{Si}[\thicknm2]$ tri-layer. Arrays of square
\microsquids with edge $\thickmicron{1\times1}$ were patterned by e-beam lithography and
$\mathrm{SF}_6$ reactive ion etching of the tri-layer. The oxidized Al layer prevents
ferromagnetic-superconductor proximity effects between the dots and the \microsquids. Although the
dots are randomly distributed on the surface, their large number yields a significant probability
to find one suitably coupled to a \microsquid. The location and shape of the single dots under
investigation were checked \aposteriori\ by AFM. The size of the dot selected
here\bracketsubfigref{fig-afm-loop-polarplots}a is $\thicknm{420\times 200\times
30}$\bracketsubfigref{fig-afm-loop-polarplots}b. Micromagnetic simulations were performed for
$\tempK0$ (no thermal activation) using custom-developed codes, either based on integrating the
LLG equation in a finite differences code (rectangular prisms)\cite{bib-TOU02} or on energy
minimization in a finite elements code (tetrahedra)\cite{bib-HER01}. The applied field was
increased step-wise in hysteresis loops. In finite differences the sample was divided into cells
with uniform lateral and vertical size $\Delta_x=\Delta_y=\unit[4.70]{nm}$ and
$\Delta_z=\unit[3.75]{nm}$, respectively. For finite elements $83310$ tetrahedra of irregular but
similar shape were used, with a maximum~(resp. minimum) volume of $\unit[42.29]{nm^3}$ (resp.
$\unit[12.50]{nm^3}$). We set $K_1=\unit[\scientific{4.8}{4}]{J.m^{-3}}$,
$A=\unit[\scientific{2}{-11}]{J.m^{-1}}$ and $\Ms=\unit[\scientific{1.73}{-6}]{A.m^{-1}}$ in the
calculation.

In the following we call $\varphi$ the angle between the in-plane $\Hext$ and the in-plane long
axis of the dot $[001]$\bracketsubfigref{fig-afm-loop-polarplots}b. Due to a shape effect,
in-plane $[1\overline10]$~($\varphi=\unit[90]{\deg}$) is a magnetically-harder direction than
$[001]$. The insets of \figref{fig-afm-loop-polarplots}c show \microsquid hysteresis loops for two
angles~($\varphi=\unit[+6;+90]{\deg}$). Such loops with negligible remanence although with
significant hysteresis, are characteristic of multidomain systems with a limited number of
domains. Starting from positive saturation the first $\Hsw$, named hereafter $\Hn$, is expected to
reveal a \emph{nucleation} event, \eg the entry of a magnetic vortex\cite{bib-GUS01b} in the dot.
The second $\Hsw$, occurring at negative fields and named $\Han$, is expected to reveal an
\emph{annihilation} event, \ie the expulsion from the dot of a previously-nucleated vortex or
wall. \figref{fig-afm-loop-polarplots}c shows the experimental angular variation $\Hn(\varphi)$
and $\Han(\varphi)$. The two-fold symmetry results from the elongated shape of the dot. Two
striking features are observed, that shall be explained in the course of the discussion. First,
jumps of both $\Hn$ and $\Han$ occur at some angles. Second, depending on the range of angles, one
or two $\Hn$ and/or $\Han$ are observed.

\begin{figure}[h]
  \includegraphics{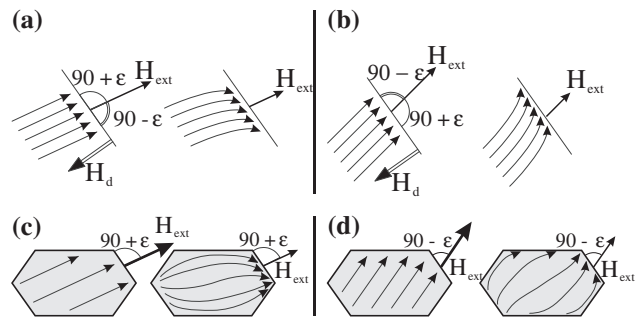}%
  \caption{\label{fig-bifurcation}\dataref{none}Schematic illustration of magnetization state bifurcation
  related to the orientation of the external field $\Hext$ with respect to an edge. Arrows length
  sketch the magnitude of $\Hext$.
  The external field $\Hext$ is tilted (a)~clockwise or (b)~anticlockwise
  with respect to the normal to the edge. For a hexagonal dot, just after bifurcation the magnetization
  state may thus be in a leaf state\cite{bib-COW98b} oriented either along the (c)~long edge or (d)~a diagonal. }
\end{figure}

\begin{figure}[h]
  \includegraphics{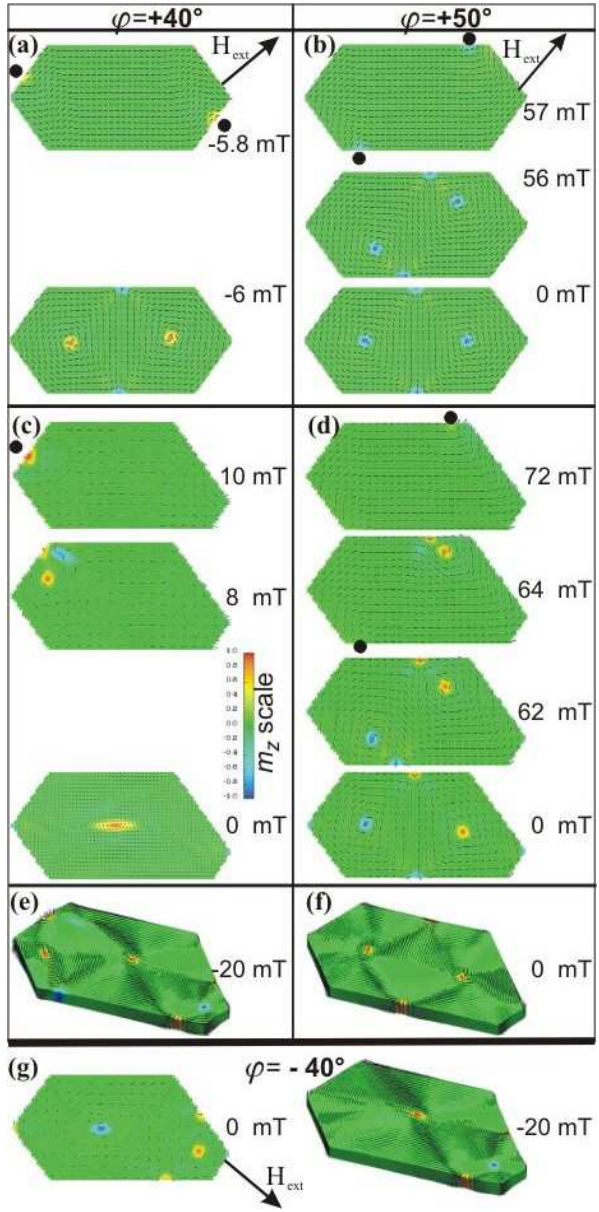}%
  \caption{\label{fig-simulation-two-angles}\dataref{none}(a-d)~Mid-height views of
  micromagnetic finite differences simulations
  of dots with vertical facets, for $\varphi=\unit[40]{\deg}$~(left) and
  $\varphi=\unit[50]{\deg}$~(right), for (a-b)~symmetric and
  (c-d)~asymmetric in-plane shape. $\Hext$ is decreased from positive saturation towards
  the flux-closure state. Nucleation loci are indicated with full dots (e-f)~Flux-closure state obtained on the same element
  with finite elements simulations~(surface views)
   (g)~Flux-closure states for $\varphi=\unit[-40]{\deg}$, for finite
  differences~(left) and finite elements~(right). The color codes the perpendicular component of magnetization~(see scale).}
\end{figure}

The jumps of $\Hsw$ can be understood qualitatively by simple arguments. Let us sketch in a
quasistatic picture the evolution of magnetization $\vect M(\vect r)$ close to an edge during the
first stages of a hysteresis loop\bracketsubfigref{fig-bifurcation}{a-b}. Starting from
saturation, upon decrease of $\Hext$ the \emph{relative} importance of the dipolar energy $\Ed$
increases. As a result $\vect M$ progressively rotates towards the edge to reduce surface charges,
and thus reduce~$\Ed$. The direction of rotation, clockwise or anticlockwise, depends on the
initial direction of $\vect M$ with respect to the normal to the edge~(imposed by the direction of
$\Hext$), due to the torque exerted by the dipolar field~$\Hd$ on $\vect{M}$. With this picture at
least two different slightly inhomogeneous magnetization states, so-called 'leaf
states'\cite{bib-COW98b}, are expected to appear upon decrease of $\Hext$, \eg when starting from
saturation along $\varphi=\unit[0]{\deg}$ or
$\varphi=\unit[90]{\deg}$\bracketsubfigref{fig-bifurcation}{c-d}. Bifurcation must occur for at
least one intermediate angle between these two paths. Then, it is obvious that for $\Hext$ applied
on either side of this angle, the magnetization pattern will evolve towards very different states,
each characterized by a different entry point for vortices--and thus of edge orientation,
explaining a jump in~$\Hn(\varphi)$. These ideas were confirmed by micromagnetic simulation. In a
first attempt the simulations were performed on a dot with vertical facets and symmetric ends.
\subfigref{fig-simulation-two-angles}{a-b} shows the static magnetization states just before and
after $\Hn$. For $\varphi=\unit[40]{\deg}$, before $\Hn$ the state belongs to the class sketched
in \subfigref{fig-bifurcation}c, as expected. Then two regions of strongly non-uniform
magnetization develop simultaneously, ending up in the entry of two vortices at~$\Hn$. As $\Hext$
is further decreased the two vortices with opposite circulation move towards the inner part of the
dot, ending up in a diamond state\bracketsubfigref{fig-simulation-two-angles}a. For
$\varphi=\unit[50]{\deg}$ the state before nucleation looks like that in
\subfigref{fig-bifurcation}d. The loci of the entering vortices are thus modified with respect to
the above situation, explaining the jump of $\Hn$, but ending as well in a diamond state, \ie with
two vortices\bracketsubfigref{fig-simulation-two-angles}b.

The above arguments explain the jumps of $\Hn$, but fail to explain (1)~the existence of either
one or two $\Hn$ and $\Han$ for some angles\bracketsubfigref{fig-afm-loop-polarplots}c (2)~the
experimental observation of both diamond and Landau states\cite{bib-FRU03c}, \ie with two or one
wall or vortex. Indeed in the simulations for any $\varphi$ two vortices appear simultaneously on
opposite loci, as the dot was assumed to be perfectly symmetric. Due to the large dot size these
vortices interact weakly with each other, thus both enter the dot, ending up in a diamond state.
In order to refine our interpretations, we now report simulations performed on dots with a
slightly asymmetric shape, similar to that of the AFM observation of the measured
dot\bracketsubfigref{fig-afm-loop-polarplots}a. Opposite loci are no more equivalent due to the
point-reversal symmetry breaking. Two situations occur. If opposite loci are similar two vortices
still enter the dot, one slightly before the other in terms of $\Hext$, however still ending in
the diamond state~(\eg for $\varphi=\unit[+50]{\deg}, $\subfigref{fig-simulation-two-angles}d). If
opposite loci are significantly different, then one of the vortices may enter the dot at a much
higher field than the other. It then moves towards its centre, delaying and possibly preventing
the entry of a second main vortex. This ends up in a Landau state~(\eg for
$\varphi=\unit[+40]{\deg}$, \subfigref{fig-simulation-two-angles}d). Notice that the vortex may
continuously change its shape into a Bloch wall, provided that the dot is long and thick
enough\cite{bib-HER99,bib-FRU03c}. Thus, an asymmetric feature~(like shape) is necessary to
explain the experimental observation of one-wall/vortex state\cite{bib-FRU03c}. Notice also that
more vortices may appear depending on the detailed shape of the
dot\bracketsubfigref{fig-simulation-two-angles}{e,g}. \subfigref{fig-afm-loop-polarplots}c~(upper
part) shows that these simulations reproduce experimental $\Hn(\varphi)$ convincingly. As a
further step some simulations were performed in the finite-element scheme to avoid numerical
roughness on the surfaces. This yielded results quantitatively similar to the case of vertical
facets.

\begin{figure}[h]
  \includegraphics{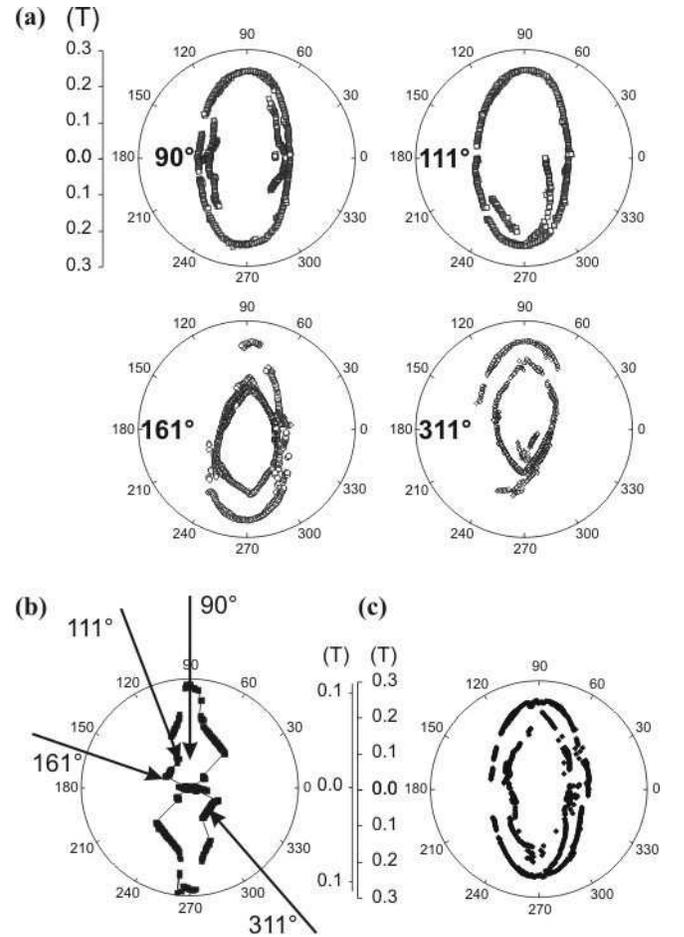}%
  \caption{\label{fig-annihilation}\dataref{none}(a)~Polar plots $\Han(\varphi_2)$
  starting from different zero-field states, each state being prepared by initial saturation
  along a given direction $\varphi_1$, shown in (b) with respect to the $\Hn(\varphi)$ plot
  (c)~Polar plot of $\Han(\varphi)$ for radial
  field sweeping~(same data as on \figref{fig-bifurcation}).}
\end{figure}

Simulations of $\Han(\varphi)$~(lower part of \subfigref{fig-afm-loop-polarplots}c) also
reproduce experimental data. The jumps of $\Han(\varphi)$ again result from the bifurcation
before nucleation, implying different states before annihilation. The occurrence of one versus
two $\Han$ may be associated with the occurrence of different flux-closure states at low field.
A more general experiment consists in proceeding to nucleation with the field decreased along a
given angle~$\varphi_1$, followed by annihilation with the field increased along a different
direction~$\varphi_2$. The plot $\Han(\varphi_2)$, measured while keeping $\varphi_1$ fixed \ie
trying to prepare the system always in the same starting state, can be viewed as a signature of
this state. \subfigref{fig-annihilation}{a} displays such plots for several values of
$\varphi_1$, chosen in different branches of the
experimental~$\Hn$\bracketsubfigref{fig-annihilation}{b}. The plots are not identical, which
confirms that the remanent state depends on the angle of $\Hext$. The fact that different plots
roughly consist of different parts of a common set of two branches is also easily understood.
For a given $\varphi_2$ a wall or a vortex of given circulation will always be pushed towards
the same locus of the dot, be it alone at remanence (vortex or Landau state) or having a
companion (diamond state). In the latter case one point is found on each branch, whereas in the
former case only one branch is revealed.

Finally, the values of $\Hsw$ depend only weakly on the algorithm used~(finite differences or
finite elements). However, sometimes fine differences appear upon nucleation, such as the
magnetization direction of vortices' cores, or the occurrence of more than two vortices~(see
\figref{fig-simulation-two-angles}, c versus e; d versus f; g), although both codes were
benchmarked successfully one against each other on time-resolved magnetization reversal
issues\cite{bib-NIST4}. This underlines that describing the fine details of nucleation with
simulations remains a challenge and that results should still be taken with care.

To conclude, we reported the measurements, quantitative reproduction and understanding of angular
nucleation and annihilation fields $\Hn(\varphi)$ and $\Han(\varphi)$ in a multi-domain magnetic
particle. This is an example of a generalization to multi-domain states of the well-known
Stoner-Wohlfarth astroid. The main feature is the occurrence of jumps in both plots, which result
from the interplay of a non-uniform magnetization state with the shape of the dot. Thus, such
jumps should not be automatically ascribed to defects when observed in experiments.

We thank Ph. David and V. Santonacci for technical support, and C. Meyer and W. Wulfhekel for
critically reading the manuscript.

\bibliographystyle{prsty}

\begin{thebibliography}{10}

\bibitem{bib-STO48}
E.~C. Stoner and E.~P. Wohlfarth, Phil. Trans. Royal Soc. London {\bf A240},
  599  (1948).

\bibitem{bib-SLO56}
J.~C. Slonczewski, Research Memo RM~003.111.224, IBM Research Center,
  Poughkeepsie, NY (unpublished).

\bibitem{bib-WER97}
W. Wernsdorfer, E.~B. Orozco, K. Hasselbach, A. Benoit, B. Barbara, N. Demoncy,
  A. Loiseau, H. Pascard, and D. Mailly, Phys. Rev. Lett. {\bf 78},  1791
  (1997).

\bibitem{bib-BON99}
E. {Bonet-Orozco}, W. Wernsdorfer, B. Barbara, A. Benoit, D. Mailly, and A.
  Thiaville, Phys. Rev. Lett. {\bf 83},  4188  (1999).

\bibitem{bib-SCH88}
M.~A. Schabes and H.~N. Bertram, J. Appl. Phys. {\bf 64},  1347  (1988).

\bibitem{bib-COW98b}
R.~P. Cowburn and M.~E. Welland, Phys. Rev. B {\bf 58},  9217  (1998).

\bibitem{bib-COW98c}
R.~P. Cowburn, A.~O. Adeyeye, and M.~E. Welland, Phys. Rev. Lett. {\bf 81},
  5414  (1998).

\bibitem{bib-HUB98b}
A. Hubert and R. Sch{\"a}fer, {\em Magnetic domains. The analysis of magnetic
  microstructures} (Springer, Berlin, 1999).

\bibitem{bib-WER96c}
W. Wernsdorfer, K. Hasselbach, A. Sulpice, A. Benoit, J.-E. Wegrowe, L. Thomas,
  B. Barbara, and D. Mailly, Phys. Rev. B {\bf 53},  3341  (1996).

\bibitem{bib-WEI02}
Z.~H. Wei, C.~R. Chang, N.~A. Usov, M.~F. Lai, and J.~C. Wu, J. Magn. Magn.
  Mater. {\bf 239},  1  (2002).

\bibitem{bib-FRU01b}
P.-O. Jubert, O. Fruchart, and C. Meyer, Phys. Rev. B {\bf 64},  115419
  (2001).

\bibitem{bib-FRU03c}
P.~O. Jubert, J.~C. Toussaint, O. Fruchart, C. Meyer, and Y. Samson, Europhys.
  Lett. {\bf 63},  135  (2003).

\bibitem{bib-WER01}
W. Wernsdorfer,  in {\em Advances in Chemical Physics}, edited by I. Prigogine
  and S.~A. Rice (Wiley, 2001), Vol.~118.

\bibitem{bib-FRE57}
E.~H. Frei, S. Shtrikman, and D. Treves, Phys. Rev. {\bf 106},  446  (1957).

\bibitem{bib-KRO62}
H. Kronm{\"u}ller, Z. Physik {\bf 168},  478  (1962).

\bibitem{bib-TOU02}
J.~C. Toussaint, A. Marty, N. Vukadinovic, J. {Ben Youssef}, and M. Labrune,
  Comput. Mater. Sci. {\bf 24},  175  (2002).

\bibitem{bib-HER01}
R. Hertel, J. Appl. Phys. {\bf 90},  5752  (2001).

\bibitem{bib-GUS01b}
K.~Y. Guslienko and K.~L. Metlov, Phys. Rev. B {\bf 63},  100403(R)  (2001).

\bibitem{bib-HER99}
R. Hertel and H. Kronm{\"u}ller, Phys. Rev. B {\bf 60},  7366  (1999).

\bibitem{bib-NIST4}
NIST standard problem 4, see
  \texttt{http://www.ctcms.nist.gov/mumag/mumag.org.html}.

\end{thebibliography}

\end{document}